\documentclass[aps,prl,twocolumn,superscriptaddress,groupedaddress]{revtex4}
\usepackage{graphicx,epstopdf} 
\usepackage{amsmath}
\usepackage{color}
\usepackage{float}
\usepackage{soul}
\usepackage{hyperref}
\usepackage{cleveref}

\newcommand{\be}{\begin{equation}}
\newcommand{\ee}{\end{equation}}
\newcommand{\bea}{\begin{eqnarray}}
\newcommand{\eea}{\end{eqnarray}}
\newcommand{\bse}{\begin{subequations}}
\newcommand{\ese}{\end{subequations}}

\begin{document}

	\title{Robust topological Hall effect driven by tunable noncoplanar magnetic state in Mn-Pt-In inverse tetragonal Heusler alloys}

	\author{Bimalesh Giri}
	\affiliation{School of Physical Sciences, National Institute of Science Education and Research, HBNI, Jatni-752050, India}
	\author{Arif Iqbal Mallick}
	\affiliation{Department of Physics, Indian Institute of Technology Bombay, Mumbai- 400076, India}
	\author{Charanpreet Singh}
	\affiliation{School of Physical Sciences, National Institute of Science Education and Research, HBNI, Jatni-752050, India}
	\author{P. V. Prakash Madduri}
	\affiliation{School of Physical Sciences, National Institute of Science Education and Research, HBNI, Jatni-752050, India}
	\author{Fran\c{c}oise Damay}
	\affiliation{Laboratoire L\'eon Brillouin, CEA-CNRS, CEA Saclay, 91191 Gif-sur-Yvette, France}
	\author{Aftab Alam}
	\affiliation{Department of Physics, Indian Institute of Technology Bombay, Mumbai- 400076, India}
	\author{Ajaya K. Nayak}
	\email{ajaya@niser.ac.in}
	\affiliation{School of Physical Sciences, National Institute of Science Education and Research, HBNI, Jatni-752050, India}

	\date{\today}

\begin{abstract}
Manipulation of magnetic ground states by effective control of competing magnetic interactions has led to the finding of many exotic magnetic states. In this direction, the  tetragonal Heusler compounds consisting of multiple magnetic sublattices and crystal symmetry favoring chiral Dzyaloshinskii-Moriya interaction (DMI) provide an ideal base to realize non-trivial magnetic structures. Here, we present the observation of a large robust topological Hall effect (THE) in the multi-sublattice Mn$_{2-x}$PtIn Heusler magnets. The topological Hall resistivity, which originates from the non-vanishing  real space Berry curvature  in the presence of non-zero scalar spin chirality, systematically decreases with decreasing the magnitude of the canting angle of the magnetic moments at different sublattices. With help of first principle calculations, magnetic and neutron diffraction measurements, we establish that the presence of a tunable non-coplanar magnetic structure arising from the competing Heisenberg exchanges and chiral DMI from the D$_{2d}$ symmetry structure is responsible for the observed THE. The robustness of the THE with respect to the degree of non-collinearity  adds up a new degree of freedom for designing THE based spintronic devices.
 
\end{abstract}
		
		
\maketitle
		
\section{INTRODUCTION}

In recent times, the art of easy manipulation of non-collinear magnetic structures over the collinear ones has led to a drastic shift of focus on research involving next generation spintronic devices. For instance, the topologically stable non-collinear magnetic objects, e.g. skyrmions, can be engineered effectively at significantly lower current densities, thereby,  providing an efficient way to manipulate the information stored in these logic/spintronic devices \cite{3,4,5,6}. In most of the cases, a basic requirement for the stabilization of these non-collinear magnetic states is the presence of chiral magnetic interaction, the Dzyaloshinskii-Moriya interaction (DMI), that develops from the broken inversion symmetry in certain class of chiral magnets \cite{7,8,9,Jin17,10} and layered thin films \cite{11,12,13}. Owing to the presence of non-trivial topological configurations, large topological Hall effect (THE) has been reported in these systems \cite{13,14,15,Saha99,16,17}. The basis of the THE can be associated with a non-vanishing scalar spin chirality (SSC) $\chi_{ijk}=\bf S_i \cdot (\bf S_j \times \bf S_k)$, that corresponds to the solid angle $\Omega$ subtended by three spins \textbf{S$_i$}, \textbf{S$_j$} and \textbf{S$_k$} on a unit sphere.  Although measurement of THE has been extensively used to characterize topological magnetic objects, the manipulation of THE for its direct use in spintronics is never demonstrated.

For the realization of THE, the system must exhibits a non-vanishing scalar spin chirality, which  can appear in magnetic materials with non-collinear and non-coplanar spin structures \cite{18,19,20}. However, the lack of chiral magnetic symmetry can force a net vanishing THE in most of these systems. Rare examples of nonzero SSC has been reported in systems with special lattice structures such as pyrochlore, triangular lattices etc., with distinctive type of spin configurations\cite{18,TriangularTHE,Machida07}.  Recent observation of THE  in a perpendicularly magnetized system with interfacial DMI corroborates the importance of chiral magnetic interaction to achieve a non-zero THE \cite{21}. Along this direction, only limited experimental findings are reported in recent literature. Our particular interest is Mn based tetragonal Heusler materials well known for their potential use in the field of spintronics \cite{10,22,23,24}. In this family of materials, a flexible tuning of the magnetic properties, such as, magnetization, magnetic anisotropy and Curie temperature ($T_C$) can be realized by tuning the sublattice magnetic moments. In addition, the non-centrosymmetric tetragonal Mn$_2$$ YZ $ compounds (where $ Y= $Pt, Rh, Ir and Ni) crystallizing in the space group I\={4}m2  are potential candidates to host DMI that can give chiral magnetic interaction in the system \cite{10,24,25,26}.  To realize our goal we select the inverse tetragonal Heusler compound Mn$_ 2 $PtIn as a starting compound that consists of two magnetic sublattices of Mn atoms. Here, we show that a tunable topological Hall effect can  be  achieved depending upon the degree of non-collinearity of the magnetic moments in the system.

\section{RESULTS AND DISCUSSION}

\begin{figure*}[tb]
	\begin{center}
		\includegraphics[angle=0,width= 16 cm, clip=true]{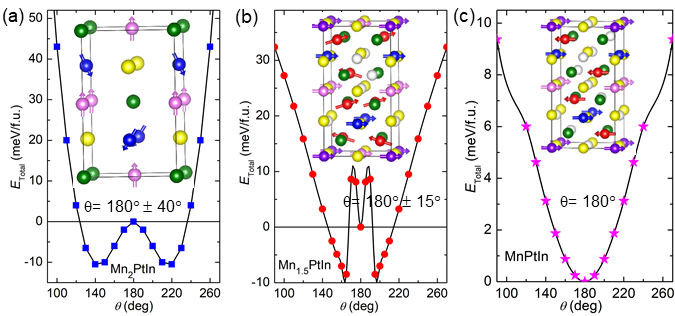}
		\caption{\label{FIG1} (Color online) First principle calculation of magnetic structure for  Mn$_{2-x}$PtIn. Total energy (with reference to E$_{\theta=180^{\circ}}$) versus canting angle $\theta$ of the Mn moment in (a) Mn$_2$PtIn, (b) Mn$_{1.5}$PtIn and (c) MnPtIn. The alignment of magnetic moment of each Mn atom in different magnetic sublattices are shown in the inset of respective figures. In case of Mn$_2$PtIn (space group; I\={4}m2), Mn occupies two positions:  Mn$_{2b}$ (magenta balls) and Mn$_{2d}$ (blue balls).  For Mn$_{1.5}$PtIn (space group; I\={4}2m) and MnPtIn (space group; I\={4}2m) the Mn sits at Mn$_{2a}$ (violet balls),  Mn$_{2b}$ (magenta balls), Mn$_{4d}$ (blue balls), Mn$_{8i}$ (red balls).  For all cases In and Pt atoms are represented by green and yellow balls, respectively.  }
	\end{center}
\end{figure*}


Calculations for different structural and magnetic configurations of the proposed Mn$_{2-x}$PtIn systems were carried out to find out the energetically most favourable crystal and magnetic configurations. In case of the parent Mn$_ 2 $PtIn, a minimum energy state was found for the tetragonal space group I\={4}m2. We have performed fixed-direction magnetic calculations using the non-collinear module of the Vienna Ab-initio simulation package (VASP). From the total energy as a function of canting angle of Mn moments between the alternate layers [Fig. 1(a)], it is found that Mn$_ 2 $PtIn exhibits a non-collinear magnetic state characterized by canting angle of 180$^{\circ}$ $\pm$40$^{\circ}$.  A maximum energy scale of about 0.35 eV/formula unit (f.u.) is the energy difference between the two collinear configurations - the ferromagnetic (FM) ($\theta$ = 0$^{\circ}$) and the ferrimagnetic (FiM) ($\theta$ = 180$^{\circ}$).  The most significant exchange coupling is between the Mn atoms in the neighboring planes, i.e. between the Mn at 2b and 2d positions.  The nearest neighbor exchange coupling aligns the Mn moments of the neighboring planes ferrimagnetically. The next nearest neighbor coupling between the Mn atoms sitting in 2d position (Mn-Pt planes) is also significantly large and it also tries to align the Mn moments antiferromagnetically. The competition between these two interactions results in an effective canting of the Mn moments at the 2d position. In case of Mn$ _{1.5} $PtIn the most stable structural configuration was obtained by utilizing the experimental lattice parameters with the space group I\={4}2m.  Mn$ _{1.5} $PtIn exhibits a non-collinear magnetic order characterized by canting angle of 180$^{\circ}$ $\pm$15$^{\circ}$ [Fig. 1(b)] and a total uncompensated moment of about 1.13$ \mu_{B} $/f.u. In case of MnPtIn, calculations were performed both for I\={4}m2 and I\={4}2m space groups. In both the cases, there can be several possible structural configurations depending on the Mn site occupancy. Irrespective of whatever magnetic configuration we start the calculation, it always stabilizes to a collinear magnetic arrangement for both the space groups [Fig. 1(c)]. A net cancellation of individual moments is achieved in case of for I\={4}m2, whereas a small uncompensated magnetic moment of 0.21 $ \mu_{B} $/f.u. is found for I\={4}2m space group.


Our experimental studies show that the parent Mn$_2 $PtIn crystallizes in an inverse Heusler tetragonal phase with space group I\={4}m2 \cite{Suppl}. Mn$_{2-x}$PtIn with $ x= $ 0.2 to 0.4 exhibit a mixed structural phases due to  composition dependent structural transition from the space group I\={4}m2 to I\={4}2m.  Samples with  $ x= $0.5 to 1.0 crystallize in the space group I\={4}2m with a small fraction of MnPt phase for $ x= $0.9 and 1.0 \cite{Suppl}. The isothermal magnetization  $M (H)$ loops measured at 2~K for different Mn$_{2-x}$PtIn samples are plotted in Fig. 2(a). The saturation magnetization intially increases for $x=$ 0 to 0.3 before   decreasing systematically with decreasing Mn concentration. As suggested by our theoretical calculations,  a fully compensated magnetic state can be found for MnPtIn that exhibit a linear kind of hysteresis loop. The variation of saturation magnetization with the Mn concentration for $ x= $ 0.0 to 1.0 is plotted in the inset of Fig. 2(a). The temperature dependence of the magnetization for the Mn$_{2-x}$PtIn are shown in the Fig. 2(b). As it can be seen, the Curie temperature ($ T_C $) systematically decreases with decreasing Mn concentration. 


\begin{figure}[tb]
	\begin{center}
		\includegraphics[angle=0,width= 8.5 cm, clip=true]{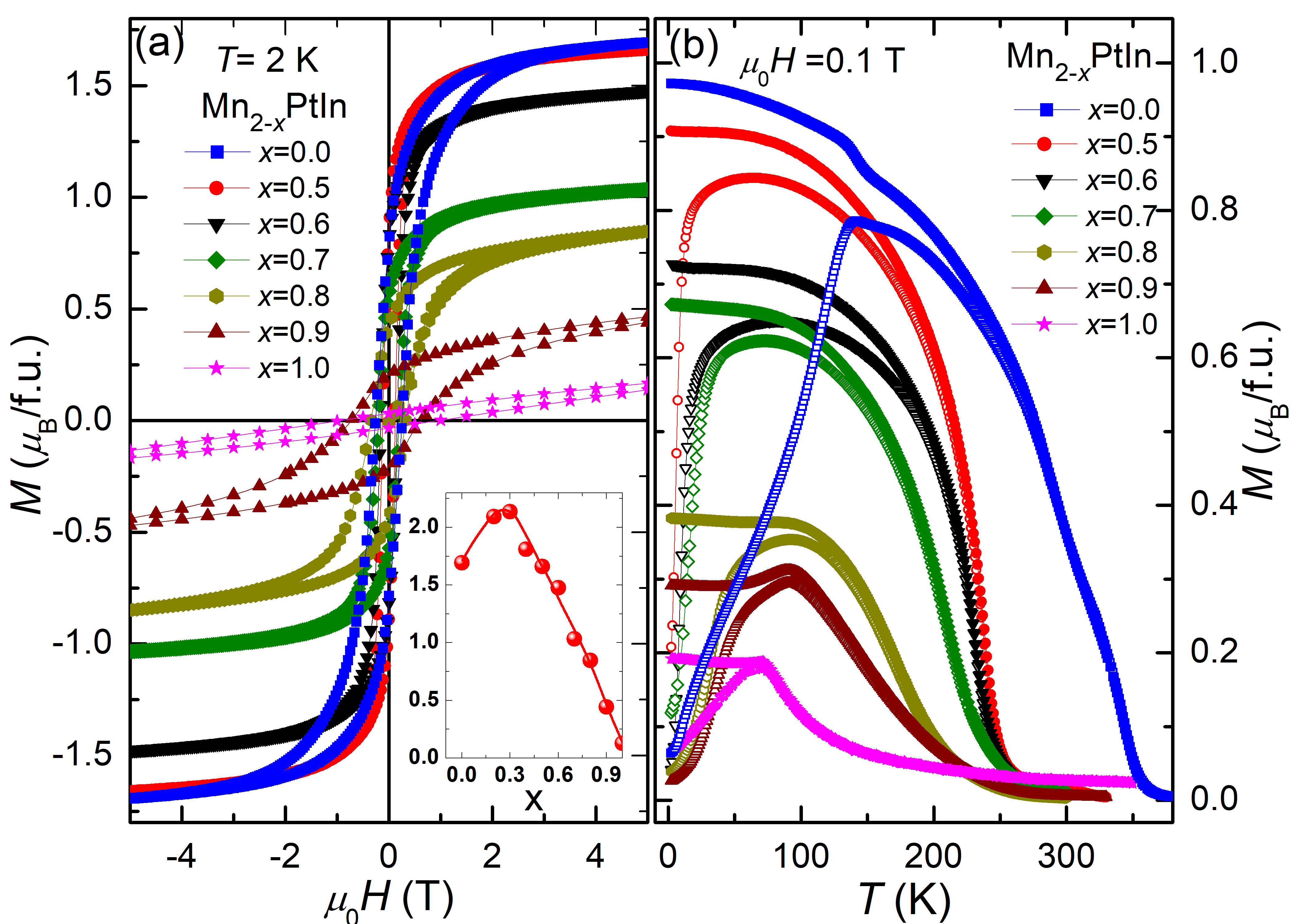}
		\caption{\label{FIG2} (Color online) (a) Field dependence of magnetization loops measured at 2~K for Mn$_{2-x}$PtIn. The inset shows compositional dependent  magnetization at a field of 5~T. (b) Temperature dependence of magnetization $ M(T) $ measured in zero field cooled (ZFC, open symbols) and field cooled (FC, closed symbols) modes in an applied field of 0.1 T for Mn$_{2-x}$PtIn. The $  M(T) $ data for x=0.9 and 1.0 are multiplied by a factor of 3 and 30, respectively, for a clear view.  }
	\end{center}
\end{figure}



\begin{figure}[tb]
	\begin{center}
		\includegraphics[angle=0,width=8.5 cm,clip=true]{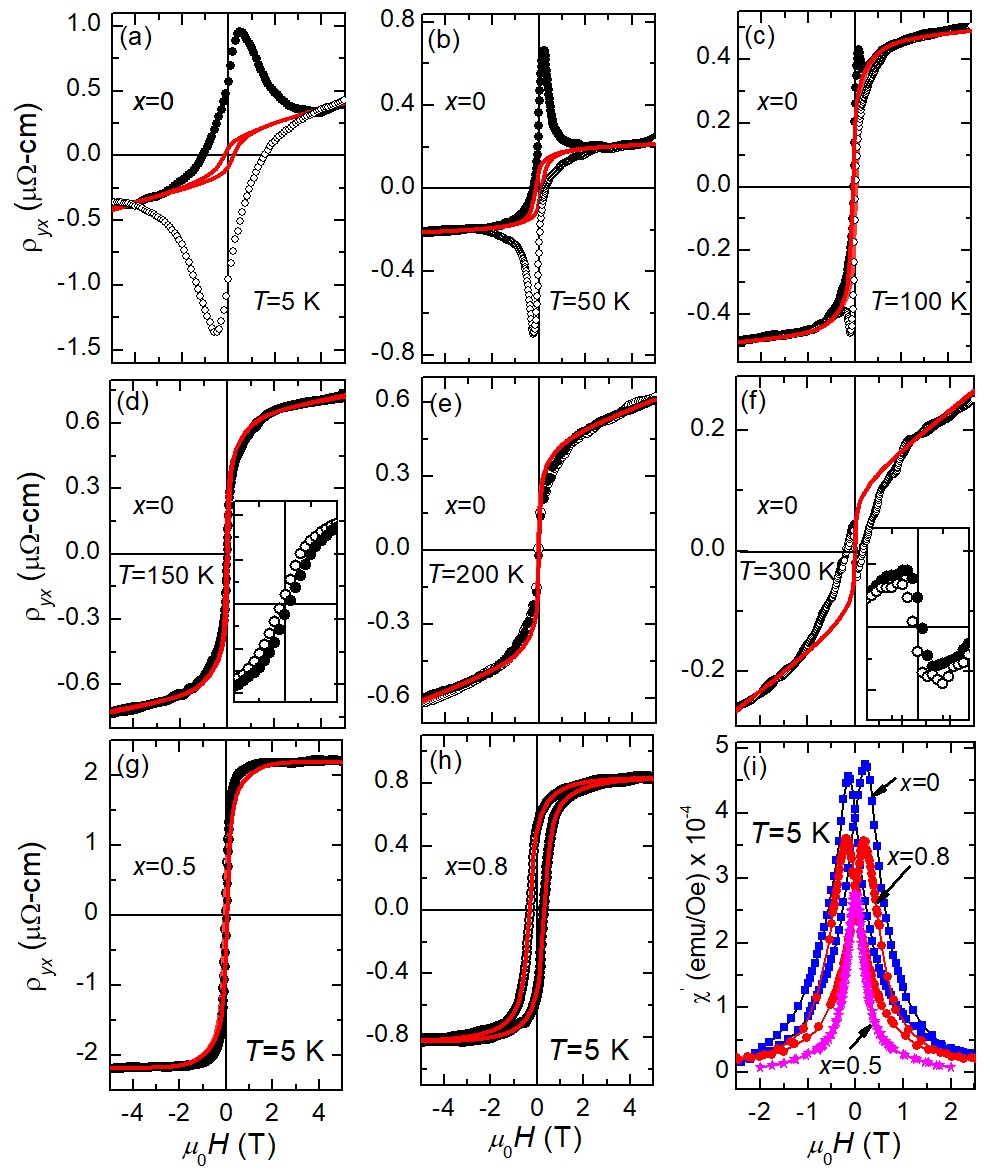}
		\caption{\label{FIG3} (Color online)  Field dependence of Hall resistivity ($\rho_{yx}$) measured at different temperatures for (a)-(f) Mn$_2$PtIn, (g) Mn$_{1.5}$PtIn and (h) Mn$_{1.2}$PtIn. The open and closed symbols represent experimental data with field sweep in $+H\rightarrow -H$ and $-H\rightarrow +H$, respectively. The solid lines corresponds to the total calculated Hall resistivity as described in the main text. (i)  Field dependence of the real component of ac-susceptibility, $\chi^{'} (H)$ measured at 5~K.  }
	\end{center}
\end{figure}
\begin{figure}[tb]
	\begin{center}
		\includegraphics[angle=0,width=8 cm,clip=true]{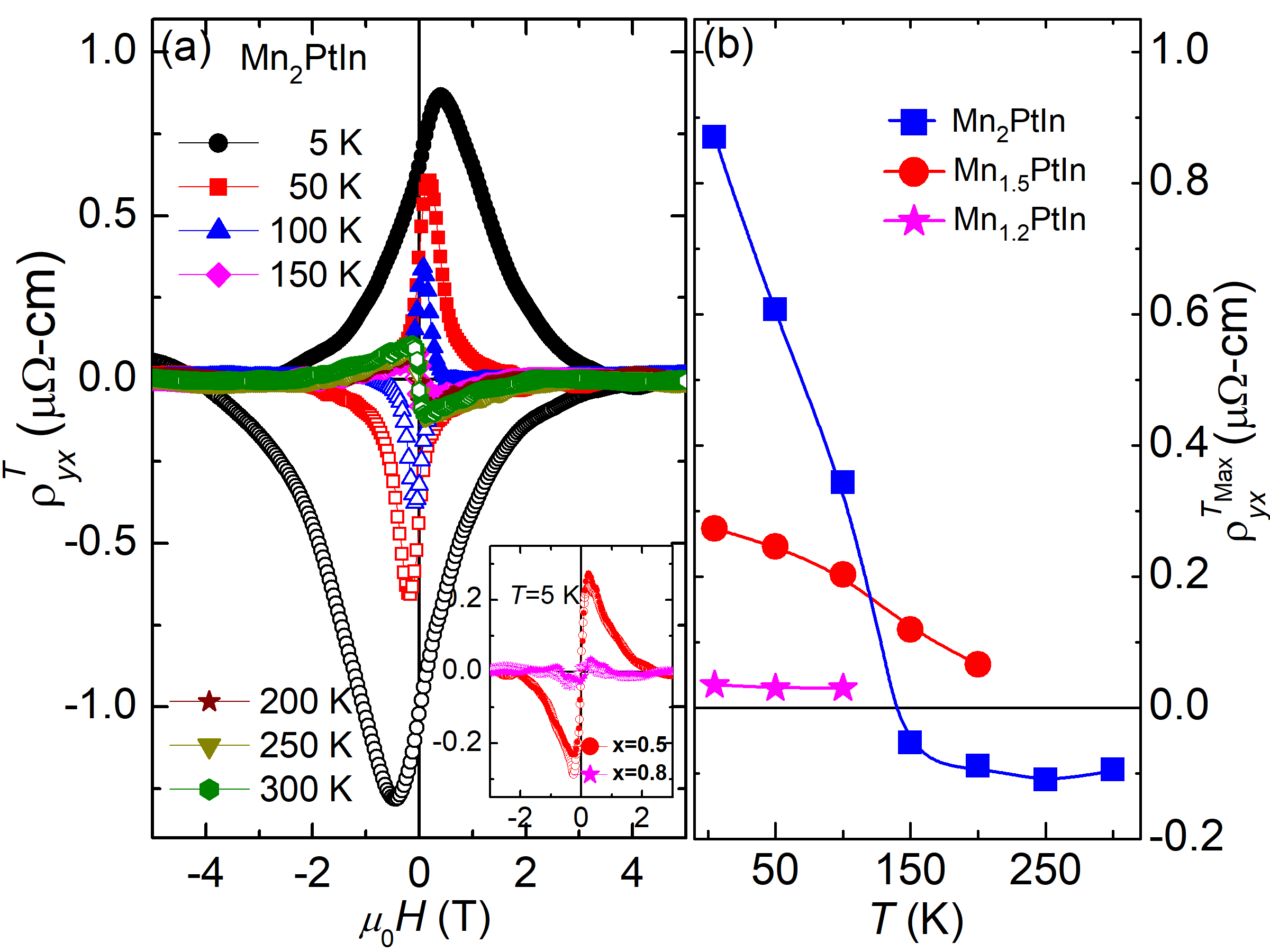}
		\caption{\label{FIG4} (Color online) (a) Topological Hall resistivity ($\rho^T_{yx}$) calculated at different temperatures for Mn$_2$PtIn. The open and closed symbols represent experimental data with field sweep in $+H\rightarrow -H$ and $-H\rightarrow +H$, respectively. The inset shows calculated $\rho^T_{yx}$  at 5~K for Mn$ _{1.5} $PtIn (filled circles) and Mn$ _{1.2} $PtIn (filled stars) in $\mu$$\Omega-$cm. (b)  Maximum value of $\rho^T_{yx}$ as a function of temperatures (solid symbols)  taken from the field dependent $\rho^T_{yx}$ data. }
	\end{center}
\end{figure}

Our theoretical calculations and experimental  studies  suggest the presence of a tunable non-collinear magnetic state in the present system. To explore the effect of this non-collinearity,  we have carried out a detailed Hall resistivity ($\rho_{yx}$) measurements for Mn$ _2 $PtIn as plotted in Fig. 3(a)-(f). It is noteworthy to mention here that an asymmetric behavior of $\rho_{yx}$ at $T=5$~K mainly arises due to small longitudinal magnetoresistance contribution at this temperature, whereas,  no such effect is found for $T\geq 50$~K. The total Hall resistivity in a system can be written as $\rho_{yx} = \rho_{N} + \rho_{AH} + \rho^T_{yx}$, where $\rho_{N}$, $\rho_{AH}$ and $\rho^T_{yx}$ are  normal, anomalous and topological Hall resistivities, respectively. The normal Hall resistivity which is linearly proportional to the magnetic field can be expressed as $\rho_{N} = R_{0}H$, where $R_0$ is the normal Hall coefficient and $ H $ is the magnetic field. In a FM/FiM system, the intrinsic contribution to the anomalous Hall effect can be illustrated  as $\rho_{AH} = b\rho_{xx}^2 M$, where $b$ is a constant, $\rho_{xx}$ is the longitudinal resistivity and $ M $ is the magnetization. As can be seen from the experimental Hall resistivity data, at higher fields $\rho_{yx}$ almost saturates with fields. So, it can be assumed that the high field data only consists of the normal and anomalous Hall components. Therefore, $\rho_{yx}$ at high fields can be written as $\rho_{yx} = R_{0}H + b\rho_{xx}^2 M$. The calculated Hall resistivity curves for different samples and temperatures sans topological Hall resistivity are plotted as solid lines in Fig. 3(a)-(h). In case of  Mn$ _{2} $PtIn, a significant difference between the experimental and calculated Hall resistivity data  can be found for temperatures up to 100~K [Figs. 3(a)-(c)], where the $\rho_{yx}$ data exhibit a negative hysteresis loop.  This suggests that for temperature range of 5~K to 100~K, the additional scattering of the conduction electrons takes place opposite to that of normal and anomalous Hall contribution. For $ T\geq 150 $~K, the  experimental $\rho_{yx}$ data exhibit positive hysteresis loop as shown in the inset of Fig. 3(d) and 3(f). We also observe a reasonable difference between the experimental and calculated Hall resistivity data for Mn$ _{1.5} $PtIn as shown in Fig. 3(g), whereas, both curves matches well for Mn$ _{1.2} $PtIn, signifying the absence of any additional component of $\rho_{yx}$ for this sample [Fig. 3(h)].

To further understand the source of the observed anomaly in the Hall data, we have performed field dependent AC-susceptibility measurements at 5~K for all the three samples [Fig. 3(i)]. The $\chi^{'} (H)$ curves do not exhibit any kind of anomaly up to a field of $\pm$5~T. It can be mentioned here that all skyrmion hosting bulk materials display dip/kink kind of features in the ac-susceptibility measurements \cite{9,24, Wilhelm11,Bauer12, Jamaluddin19}. In addition, the first derivative of magnetization with respect  to field for the $ M(H) $ loops also do not exhibit any unusual behavior \cite{Suppl}. Therefore, it is very unlikely that the present samples possess any kind of skyrmionic phase.
\begin{figure}[tb]
	\begin{center}
		\includegraphics[angle=0,width=8 cm,clip=true]{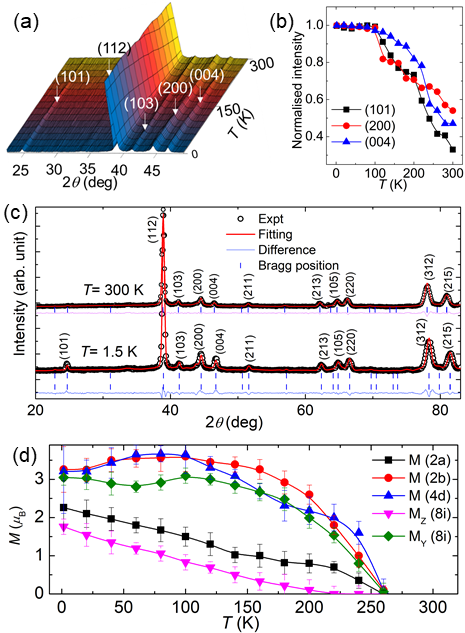}
		\caption{\label{FIG5} (Color online)   (a) Neutron diffraction patterns for Mn$_ {1.5} $PtIn measured at different temperatures. For a clear view of the major magnetic reflections the patterns are shown in the 2$ \theta $ range of 22 degree to 50 degree. (b) Temperature variation of normalized integrated intensity for the three major magnetic reflections (101), (200) and (004). (c) Rietveld refinement of the neutron diffraction patterns at 300~K and 1.5~K for Mn$_{1.5} $PtIn. (d) Temperature dependence of net magnetic moment of site-specific Mn atoms at different sublattices as depicted by different symbols. }
	\end{center}
\end{figure}

The calculated $\rho_{yx}$ (=$\rho_{N}$ + $\rho_{AH}$) was subtracted from the experimental total Hall resistivity to obtain the topological Hall  $\rho^T_{yx}$. For Mn$ _2 $PtIn, it is found that for $T \leq$100~K,  the maximum value of $\rho^T_{yx}$ lies in the first and the third quadrants, whereas,   the maxima lies in the second and fourth quadrants for $T \geq$150~K [Fig. 4(a)]. A large $\rho^T_{yx}$ of about 1~$\mu$$\Omega-$cm can be found at 5~K in Mn$ _2 $PtIn. The $\rho^T_{yx}$ drastically decreases to about 0.3~$\mu$$\Omega-$cm in case of Mn$ _{1.5} $PtIn before vanishing for Mn$ _{1.2} $PtIn, as depcted in inset of Fig. 4(a). Most importantly, $\rho^T_{yx}$ displays a strong correlation with the magnitude of canting angle in the system. To understand the chnage in sign of the $\rho^T_{yx}$ above 150~K for  Mn$ _2 $PtIn, we have plotted maximum value of $\rho^T_{yx}$ taken from the field sweep $-H\rightarrow +H$ at different temperatures [Fig. 4(b)].  The origin of the change in sign of the $\rho^T_{yx}$ can be attributed to the existence of a spin-reorientation like transition at 150~K as can be visualized in ZFC and FC $ M(T) $ curves shown earlier in Fig. 2(b). In case of Mn$ _{1.5} $PtIn the maximum value of $\rho^T_{yx}$ monotonically decreases with temperature, whereas, it remains almost close to zero at all temperatures for Mn$ _{1.2} $PtIn. It is important to mention here that the $ M(T) $ data for both Mn$ _{1.5} $PtIn and Mn$ _{1.2} $PtIn do not exhibit any kind of anomaly [see Fig. 2(b)].

\begin{figure}[tb]
	\begin{center}
		\includegraphics[angle=0,width=8 cm,clip=true]{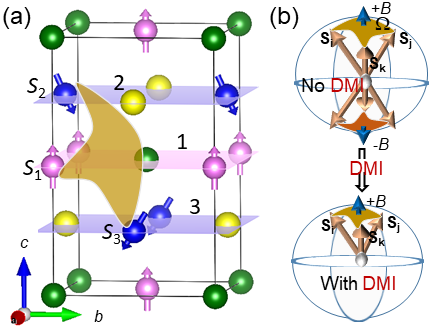}
		\caption{\label{FIG6} (Color online)    (a) Unit cell for Mn$_2$PtIn representing Mn moments at different lattice planes. Mn-In and Mn-Pt lattice planes are shown in light magenta color marked by 1 and  light blue color  marked by 2 and 3, respectively. Solid angle subtaineded by three moments $\bf S_1$, $\bf S_2$ and $\bf S_3$ is ahown in dark yellow color. (b) Upper panel; solid angle $\Omega$ subtended by three non-coplanar spins $ {\bf S_i} $, $ {\bf S_j} $, $ {\bf S_k} $ that gives a fictitious magnetic field in both upward and downward direction (blue arrows) in the absence of any chiral DMI. Lower panel; fixed chirality in the presence of DMI. }
	\end{center}
\end{figure}


Finally, to experimentally verify the existence of non-collinear magnetic structure in the present system we have performed powder neutron diffraction (ND) experiment on Mn$_{1.5} $PtIn. The ND patterns taken in the temperature range of 1.5 K to 300 K at different 2$ \theta $ values are depicted in Fig. 5(a). The temperature dependent ND patterns show an increase in the scattering intensity below the ordering temperature at the nuclear Bragg peaks (101), (200) and (004), suggesting a commensurate magnetic structure. This can be clearly seen from the temperature variation of normalized intensities which decrease significantly with increasing temperature, suggesting the presence of contribution from both in-plane and out-of-plane magnetic components [Fig. 5(b)]. Since Mn$_{1.5} $PtIn exhibits a $T_C $ of about 240~K, the ND pattern at 300~K used for the nuclear refinement by utilizing the previously determined space group I\={4}2m (SG No. 121) and related structural parameters [Fig. 5(c), upper panel]. Furthermore, we obtained the magnetic propagation vector $ k= $ (0, 0, 0) with best agreement factors by using the $ k $-search programme included in Fullprof-suite package. The Rietveld refinement of 1.5~K ND data convincingly demonstrates the presence of  magnetic contribution  in Mn$_{1.5} $PtIn [Fig. 5(c), lower panel]. The temperature dependence of absolute values of the magnetic moments for Mn sitting at different sublattices are shown in Fig. 5(d).  The Mn sitting at 2b, 4d and 8i display almost equal magnitudes of magnetic moments with similar temperature dependance. The Mn moments at 2a, 2b and 4d exhibit a complete in-plane orientation, whereas, 8i Mn atoms possess both in-plane and out-of-plane magnetic components. A smaller magnitude of 2a Mn moment is due to the fact that 2a site is comparatively less occupied in the present sample.   More details about the analysis of ND data can be found from the supplementary information \cite{Suppl}.

Our theoretical calculation and experimental results have convincingly established the presence of non-coplanar magnetic state in the present system. For a better understanding of the present THE that originates from the non-vanishing scalar spin chirality, we have considered one Mn spin ($\bf S_1$) in the Mn-In plane (plane 1) and two Mn spins ($\bf S_2$ and $\bf S_3$) from two different Mn-Pt planes (plane 2 and 3) as shown in Fig. 6(a). For simplicity, first we consider the components of the canted spins $\bf S_2$ and $\bf S_3$ lie in $y-z$ plane. The magnetic moment of Mn atoms sitting in plane 1, 2 and 3 can be described by $\bf S_1$ = $Z_1$ $\bf k$, $\bf S_2$ = $Y_2$ $\bf j$ - $Z_2$ $\bf k$, $\bf S_3$ = -$Y_2$ $\bf j$ - $Z_2$ $\bf k$, respectively. Here $Y_2$ is the component of Mn moment in the $y$ direction and $Z_1$  and $Z_2$ are that of $z$ direction. $ \bf i, j, k$ are the unit vectors. For the said configuration the scalar spin chirality can be calculated as $\chi_{123}=\bf S_1 \cdot (\bf S_2 \times \bf S_3)$= 0. However, the competing antiferromagnetic interactions along with  the chiral DMI in the system and/or the external magnetic field can tilt the in-plane component of the  $\bf S_2$ and $\bf S_3$ in any direction in the $ab$ plane. Hence, with a small $x$ component $\delta$,  $\bf S_2$= $\delta \bf i$ + $Y_2$ $\bf j$ - $Z_2$ $\bf k$ and $\bf S_3$ = $\delta \bf i$ - $Y_2$ $\bf j$ - $Z_2$ $\bf k$. As a result, we can achieve a non-vanishing $\bf S_1 \cdot (\bf S_2 \times \bf S_3)$= $-2Y_2Z_1 \delta $ that can give rise to the observed topological Hall effect. As schematically depicted in Fig. 6(b), any three non-coplanar spins $ {\bf S_i} $, $ {\bf S_j} $ and $ {\bf S_k} $ can subtend a solid angle $\Omega$, thereby resulting in a non-zero scalar spin chirality with a fictitious magnetic field as shown by blue arrow. In absence of any fixed chirality  this magnetic field will act in all possible directions resulting in a net vanishing THE. However, the $ D_{2d} $ symmetry of the present materials ensures a  chiral magnetic state, thereby, a non-vanishing THE in the system. At very high magnetic fields the Zeeman energy can easily overcome the chiral DMI energy, suppressing the chirality as well as the THE. 

Our assertion of large topological Hall resistivity at low temperatures as a result of finite scalar spin chirality instead of any skyrmion phase is supported by several facts. (i) We do not observe any kind of kink/peak behavior in the ac-susceptibility measurements as discussed earlier. (ii)  The isostructural Heusler compounds Mn$_{1.4}$Pt$_{0.9}$Pd$_{0.1}$Sn \cite{10} and Mn$_2$Rh$_{0.9}$Ir$_{0.1}$Sn  \cite{skx20}  found to exhibit antiskyrmion phase with antiskyrmion size of about 150~nm and 200~nm, respectively . Since the magnitude of THE is inversely proportional to the skyrmion size (density), it is expected that the antiskyrmion phase in these systems will result in topological Hall resistivity in the order of 1~n$\Omega-$cm or less. (iii) It can be clearly seen that Mn$_2$PtIn displays a spin-reorientation transition  around 150~K. The previous studies on similar systems show the existence of antiskyrmion phase, if any, only above the spin-reorientation transition \cite{10, Jamaluddin19}. 

The non-coplanar spin structure with finite scalar spin chirality as a source of THE has been recently observed  in FM systems \cite{21}, as well as in antiferromagnetic systems (AFM) \cite{THE2014}. Although THE arises in systems hosting skyrmions/antiskyrmions or comprised of non-coplanar spin structures, a very basic difference lies in the length scale of the periodicity associated with their magnetic structure. A crossover between a long periodic magnetic structure to a comparatively shorter scale non-coplanar magnetic state is about interplay between the energy contributions from various energy terms \cite{Bogdanov02}. In case of the present systems under study, it might be possible to stabilize a incommensurately modulated helical/cycloid ground state with a modulation period up to few hundreds. These helix/cycloid can transform into skyrmions/antiskyrmions under the external magnetic field and in turn can cause  very small THE due to the large size of the skyrmions. In a recent study, Kumar $ et $ $ al. $ have assigned the low temperature THE in some of their samples to the presence of antiskyrmions, although these samples display a large canting angle below the spin-reorientation transition \cite{Kumar20}. However, the THE is only found when there is a large canting angle, indicating its probable origin from the non-vanishing scalar spin chirality coming from the non-coplanar magnetic state. This scenario is supported by non existence of any THE in the well-established antiskyrmion phase, probably due to the large size of the antiskyrmions in these materials.

 \section{CONCLUSION}

In summary, our theoretical calculations as well as exprimental findings  convincingly establish the presence of non-collinear magnetic ground states,  resulting in a robust THE. We show a controlled tuning of the topological Hall effect by modifying the canted magnetic state in the  system. The magnitude of the spin canting achieved by tuning the Mn composition is associated with a small change in lattice parameters that eventually controls  various fundamental parameters such as the exchange interactions, the DMI, and the magnetocrystalline anisotropy (MCA). Hence, the competition among these parameters determines the underlying magnetic texture of the present system.  Therefore, the THE in the present case can be controlled electrically by inducing strain in the system. Hence, the present study on realization of tunable THE possesses a great potential in all electrical switching based memory application. A very recent study on electrical control of anomalous Hall state corroborate the importance of the present study that can motivate further research in this direction \cite{Tsai20}.


\begin{acknowledgments}
	
AKN acknowledges the support from Department of Atomic Energy (DAE), the Department of Science and Technology (DST)-Ramanujan research grant (No. SB/S2/RJN-081/2016), SERB research grant (ECR/2017/000854) and Nanomission research grant  [SR/NM/NS-1036/2017(G)] of the Government of India. AA thank IRCC, IIT Bombay for supporting this research through Early Carrier Award grant, Code: RI/0217/10001338-001.

\end{acknowledgments}


\pagebreak
\section{SUPPLEMENTARY INFORMATION}

\section{Methods}
Polycrystalline ingots of Mn$ _{2-x} $PtIn with  x= 0.0 to1.0 were prepared by arc melting technique using the stoichiometric amounts of high purity constituent elements in a high purity argon atmosphere.  To ensure a better homogeneity, the samples were melted multiple times by flipping upside down. The ingots were sealed in an evacuated quartz tube and subsequently annealed at 1023 K for seven days followed by quenching in ice-water mixture. The structural phase of the annealed samples was determined by room temperature powder X-ray diffraction (XRD) using Cu-K$_{\alpha}$  radiation (Rigaku). The homogeneity and composition of the samples were determined by field emission scanning electron microscopy (FESEM), energy-dispersive X-ray spectroscopy (EDS). Magnetic measurements were carried out using vibrating sample SQUID magnetometer (MPMS-3, Quantum Design). Transport measurements were carried out using Quantum Design Physical Property Measurement System (PPMS). Powder neutron diffraction measurements were performed in Laboratoire L\'eon Brillouin-Orphee, CEA-CNRS Saclay, using G4.1 set up with neutron wavelength of 2.426 \AA.
Theoretical calculations were performed using density functional theory (DFT) as implemented within the Vienna ab-initio simulation package (VASP) \cite{supplement1}, with a projected augmented basis \cite{supplement2}. The exchange-correlation functionals used for the noncollinear magnetic calculations are based on the local density approximation (LDA). A reasonably high plane-wave cut-off of 400 eV and a Monkhorst-Pack k -point mesh with reasonable density for different unit cells were used for the calculations.
\begin{figure}[tb]
	\begin{center}
		\includegraphics[angle=0,width=8 cm,clip=true]{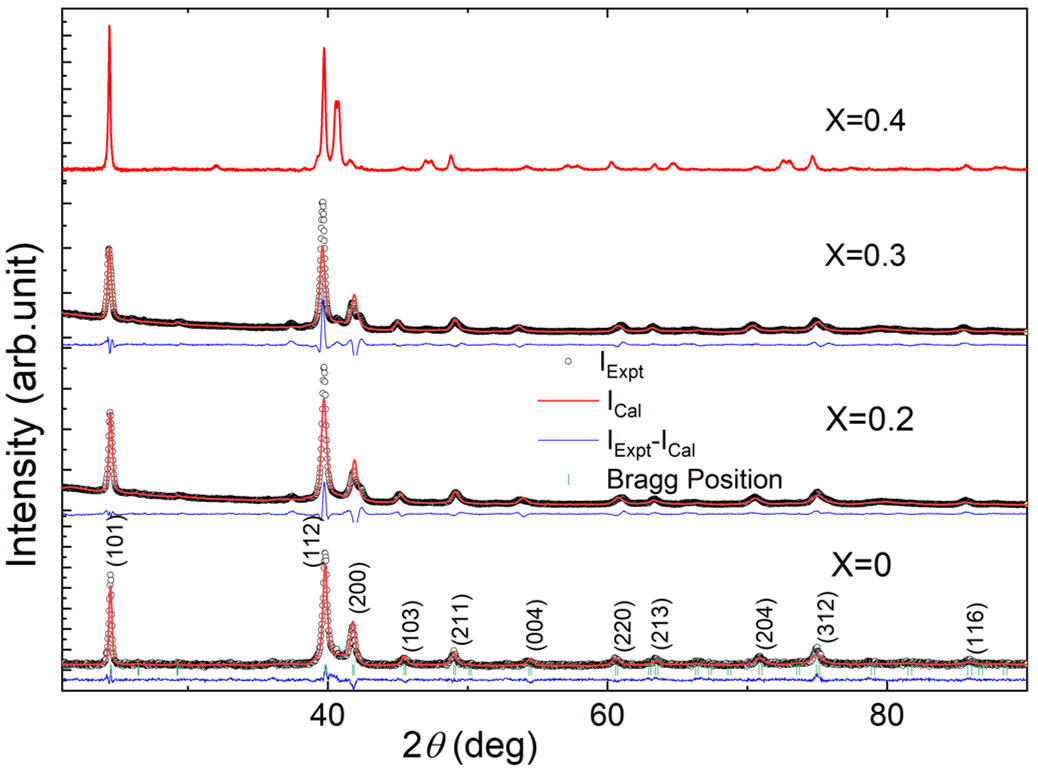}
		\caption{\label{FIG7} (Color online)   Room temperature XRD patterns for Mn$ _{2-x} $PtIn for x =0.0 to 0.4. The open black circles represent the experimental data, the red color fitted lines to the experimental data represent the simulation pattern obtained from the Rietveld refinement. The difference between the experimental and simulation patterns is shown below the respective XRD patterns. The scattered vertical lines indicates Bragg's positions. The (h k l) value for all major reflections are shown for x =0.0. }
	\end{center}
\end{figure}

\begin{figure}[tb]
	\begin{center}
		\includegraphics[angle=0,width=8 cm,clip=true]{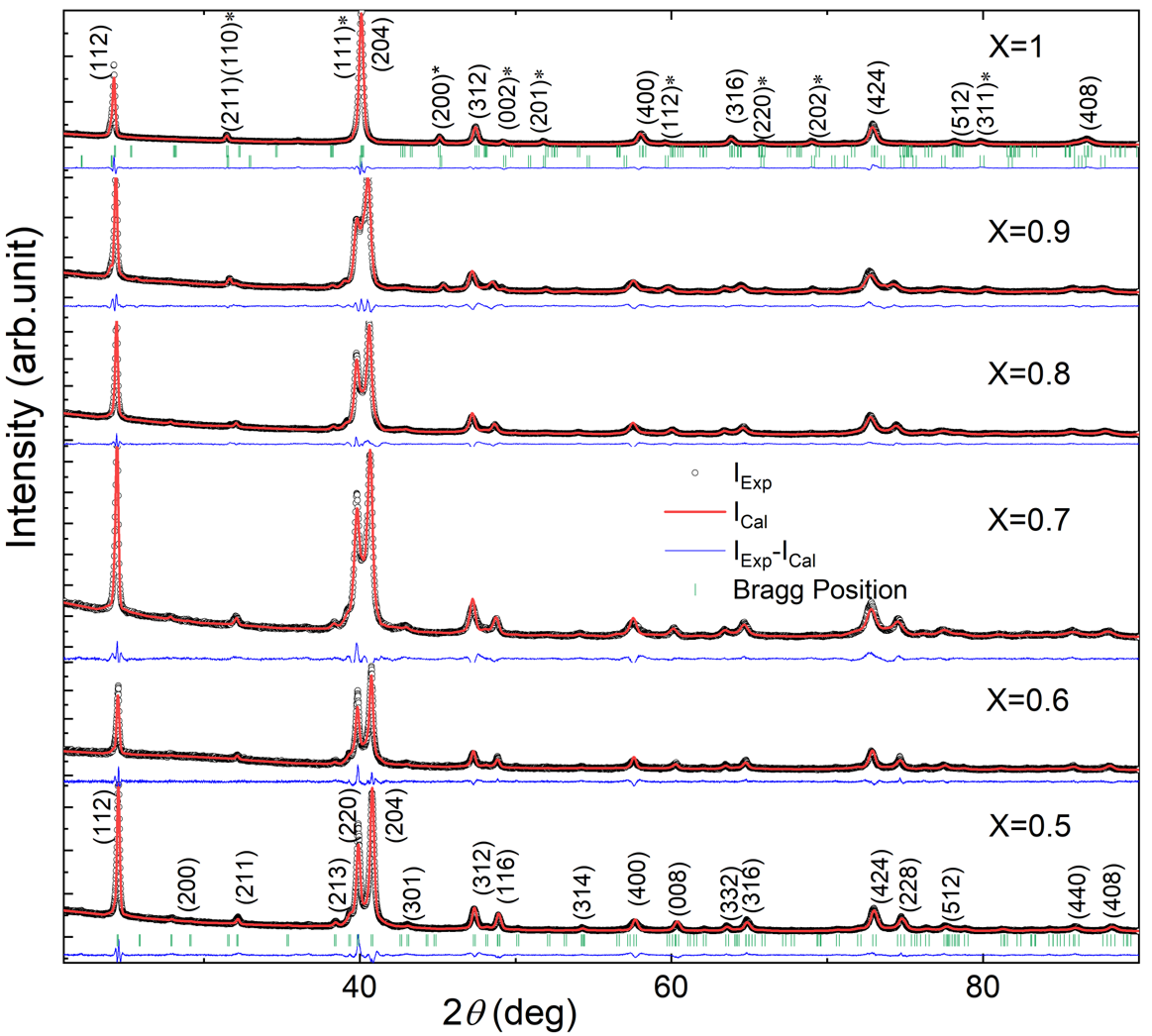}
		\caption{\label{FIG8} (Color online)   Room temperature XRD patterns for Mn$ _{2-x} $PtIn for x =0.5 to 1.0.  The open black circles represent the experimental data and the red color fitted lines to the experimental data after Rietveld refinement. The difference between the experimental and simulation patterns is shown below the respective XRD patterns. The scattered vertical lines indicates Bragg's positions. The (h k l) value for all major reflections are shown for x =0.5 and 1.0. The ($ \star $) marked (h k l) indexes for x =1.0 belongs to the MnPt secondary phase.}
	\end{center}
\end{figure}
\section{Structural Analysis} 
\begin{figure}[tb]
	\begin{center}
		\includegraphics[angle=0,width=8 cm,clip=true]{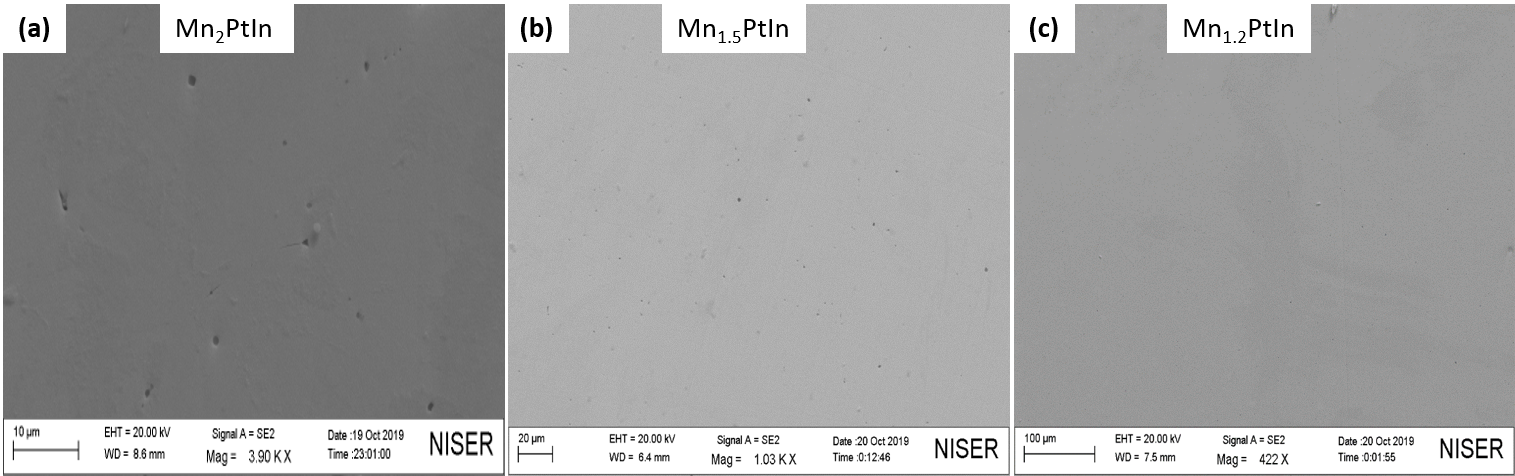}
		\caption{\label{FIG9} (Color online)  SEM images of (a) Mn$ _{2} $PtIn, (b) Mn$ _{1.5} $PtIn and (c) Mn$ _{1.2} $PtIn.}
	\end{center}
\end{figure}
\subsection{XRD Analysis} 
Room temperature powder XRD patterns with their corresponding Rietveld refinement for Mn$ _{2-x} $PtIn with x= 0.0 to 1.0 are presented in Fig. 7 and Fig. 8. We have performed Rietveld refinement with the space group (SG) I\={4}m2 (SG No. 119) for x=0.0 to 0.3  and I\={4}2m (SG No. 121) for x=0.5 to 1.0. The parent compound Mn$ _{2}$PtIn crystallizes in the inverse Heusler tetragonal structure which belongs to the space group I\={4}m2. In this structure the Mn atoms sits at the Wyckoff positions 2b (0, 0, 0.5), and 2d (0, 0.5, 0.75), whereas, In and Pt atoms occupies the 2a (0, 0, 0) and 2c (0, 0.5, 0.25), respectively. Samples with composition x=0.2 to 0.4 show a structurally mixed-phase due to the composition-dependent structural transition from SG I\={4}m2 to I\={4}2m. Samples with the composition ranges x=0.5 to 1.0, also crystallize in tetragonal structure and belong to the space group I\={4}2m (SG No. 121). Within this structure the Mn atoms occupy four different Wyckoff positions, i.e. Mn$ _{I} $ at 2a (0, 0, 0), Mn$ _{II} $ at 2b (0, 0, 0.5), Mn$ _{III} $ at 4d (0, 0.5, 0.25) and Mn$ _{IV}  $at 8i (x, x, z). Pt atoms occupies two Wyckoff sites 4c (0, 0.5, 0) and 4e (0, 0, 0.26473), whereas, In sits at the 8i (x$ ^{'} $,  x$ ^{'} $,z$ ^{'} $). Here it can be mentioned that a small fraction of MnPt phase (CuAu- I type tetragonal structure) \cite{supplement3} is observed for x =0.9 and 1.0 samples. The MnPt phase fractions were estimated to be approximately 13\% and 16\% in the case of Mn$ _{1.1} $PtIn and MnPtIn respectively. Since MnPt shows an antiferromagnetic \cite{supplement3} ordering with zero net magnetic moments, it should not affect the magnetic properties of Mn$ _{1.1} $PtIn and MnPtIn samples. The structural and magnetic parameters of Mn$ _{2-x} $PtIn are summarized in Table-II.
\begin{table*}[tb]
	\caption{Comparison of exact stoichiometric atomic percentages with obtain data from EDS}
	\centering
	\begin{tabular}{p{0.25\linewidth}p{0.25\linewidth}p{0.25\linewidth}}
		\hline
		Selected Composition &Exact atomic ratio (\%) & Obtained from EDS (\%)  \\
		\hline
		Mn$_{2}$PtIn & Mn-50, Pt-25, In-25 & Mn-52.03, Pt-25.62, In-22.34 \\
		Mn$_{1.5}$PtIn & Mn-42.85,	Pt- 28.57, In-28.57 & Mn-42.53, Pt-29.38, In-28.06  \\
		Mn$_{1.2}$PtIn &Mn-37.50, Pt-31.25, In-31.25 & Mn-40.74, Pt-30.45, In-28.80 \\
		\hline
	\end{tabular}
\end{table*}
\subsection{SEM and EDS Analysis}
The compositional homogeneity in the present samples were determined with help of Scanning Electron Microscopy (SEM). The SEM images for three samples that are mostly used for topological Hall effect measurement are shown in Fig. 9. The single-phase nature of the samples can be clearly seen from the SEM images. Small black spots in the Fig. 9(a) is due to the presence of small holes on the sample surface. To know the chemical composition of these samples we have performed energy dispersive X-ray analysis (EDS) at several places on the sample. As it can be seen from Table- I, the chemical composition obtained from the EDS analysis nearly matches with the starting composition of the samples. 
\begin{table*}[tb]
	\caption{Structural and magnetic parameters}
	\centering
	\begin{tabular}{p{0.13\linewidth}p{0.13\linewidth}p{0.13\linewidth}p{0.13\linewidth}p{0.13\linewidth}p{0.13\linewidth}p{0.13\linewidth}}
		\hline
		Composition(x) &structure  (Space Group) &  $a=b$ ( in $\AA$) & $c$  ( in $\AA$)& c/a ratio & $T_C $ (K) &$  M_S  $ ($\mu_{B}$/f.u.)   Experimental (at 5T field) \\
		\hline
		x= 0.0 & $ Tet. $ (I\={4}m2) & 4.32(1) & 6.74(3)& 1.56 &344  &1.69  \\
		x= 0.2 & $ Tet. $ (I\={4}m2) & 4.30(8)  & 6.79(0)& 1.57 &305  & 2.09 \\
		x= 0.3 & $ Tet. $ (I\={4}m2)$^*$ & 4.30(7) & 6.82(4)& 1.58 & 314 & 2.13 \\
		x= 0.4 & $ Tet. $ (I\={4}m2)$^*$ & 4.42(8)  & 6.32(4)& 1.43 & 325 &  1.81 \\
		x= 0.5 & $ Tet. $ (I\={4}2m) & 6.39(6) & 12.27(0)&1.91 & 239 & 1.66 \\
		x= 0.6 & $ Tet. $ (I\={4}2m) & 6.40(1) &12.28(2)& 1.92 & 231  & 1.47 \\
		x= 0.7 & $ Tet. $ (I\={4}2m) & 6.40(6)  &12.30(7)& 1.92 &  208 & 1.03\\
		x= 0.8 & $ Tet. $ (I\={4}2m) & 6.40(8) & 12.33(0)& 1.92 & 172 & 0.84 \\
		x= 0.9 & $ Tet. $ (I\={4}2m)$^\dagger$ & 6.40(3)  &12.37(1)& 1.93 & 141 &  0.44 \\
		x= 1.0 & $ Tet. $ (I\={4}2m)$^\dagger$ & 6.36(0) &12.67(2)& 1.99 & 82 &  0.13 \\
		\hline
	\end{tabular}
	
	$*$ siginifies a structuraly mixed phase.
	$\dagger$ represents the presence of secondary MnPt phase. 
\end{table*}
\section{Magnetic Studies}
The temperature dependence of the magnetization for the Mn$ _{2-x} $PtIn for x=.0 and x = 0.5 to 1.0 are shown in the main manuscript. As it can be seen, the Curie temperature systematically decreases with decreasing Mn concentration. We have summarized various parameters related to the structural and magnetic properties of the present samples in Table-II. Fig. 10(a)-(c) represents the first derivative of the M-H data taken at 5K for Mn$ _{2} $PtIn, Mn$ _{1.5} $PtIn and Mn$ _{1.2} $PtIn respectively.  The DC magnetic field dependent AC susceptibility data of these samples also show similar behavior. Fig. 10(d) shows the second derivative of the magnetization with respect to the magnetic field. This gives the field at which the magnetization loop changes its slope just before going to a single domain state. The purpose is to compare the field obtained from the second derivative of magnetization with the field where we observe a maximum topological Hall effect.
\begin{figure}[tb]
	\begin{center}
		\includegraphics[angle=0,width=8 cm,clip=true]{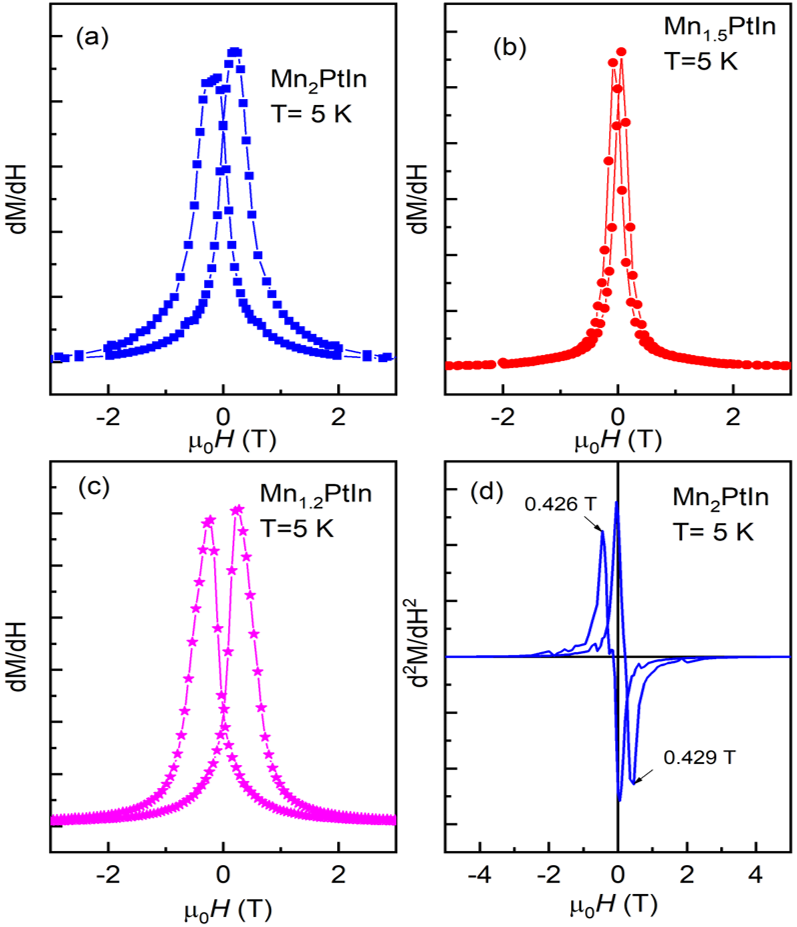}
		\caption{\label{FIG10} (Color online)  .  The first derivative of the M-H data taken at 5K for (a) Mn$_{2}$PtIn, (b) Mn$ _{1.5} $PtIn and (c) Mn$ _{1.2} $PtIn. (d) The second derivative of the M-H data taken at 5K for Mn$_{2}$PtIn.}
	\end{center}
\end{figure}
\section{Powder Neutron Diffraction}
\begin{figure}[tb]
	\begin{center}
		\includegraphics[angle=0,width=8 cm,clip=true]{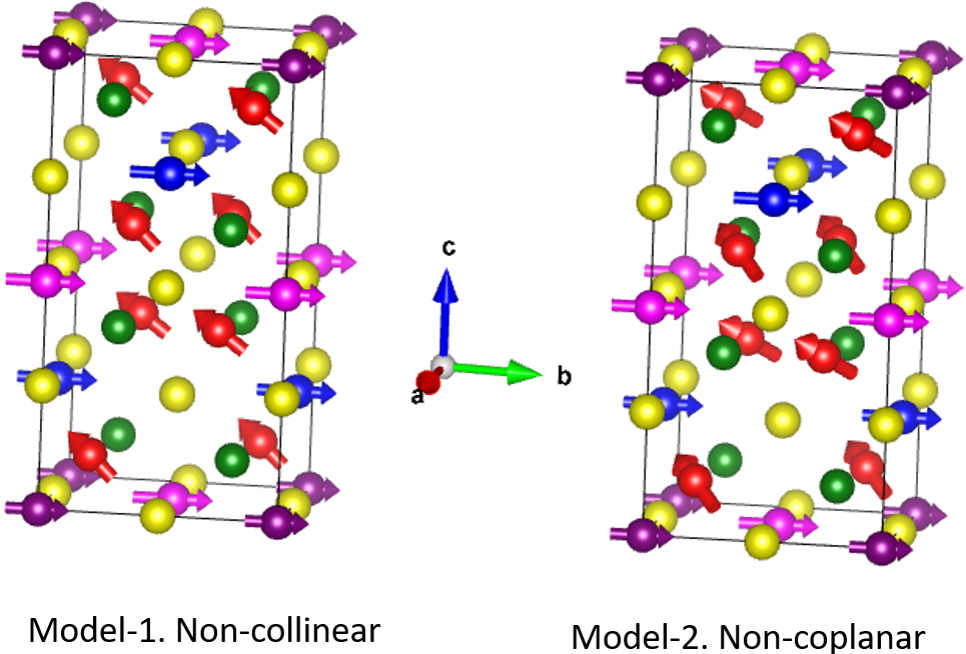}
		\caption{\label{FIG11} (Color online) Magnetic structure for two different model of Mn$ _{1.5} $PtIn at 1.5 K. For the visual clarity the magnitude and direction of Mn(8i) moments are not at per scale. }
	\end{center}
\end{figure}


\begin{figure}[tb]
	\begin{center}
		\includegraphics[angle=0,width=8 cm,clip=true]{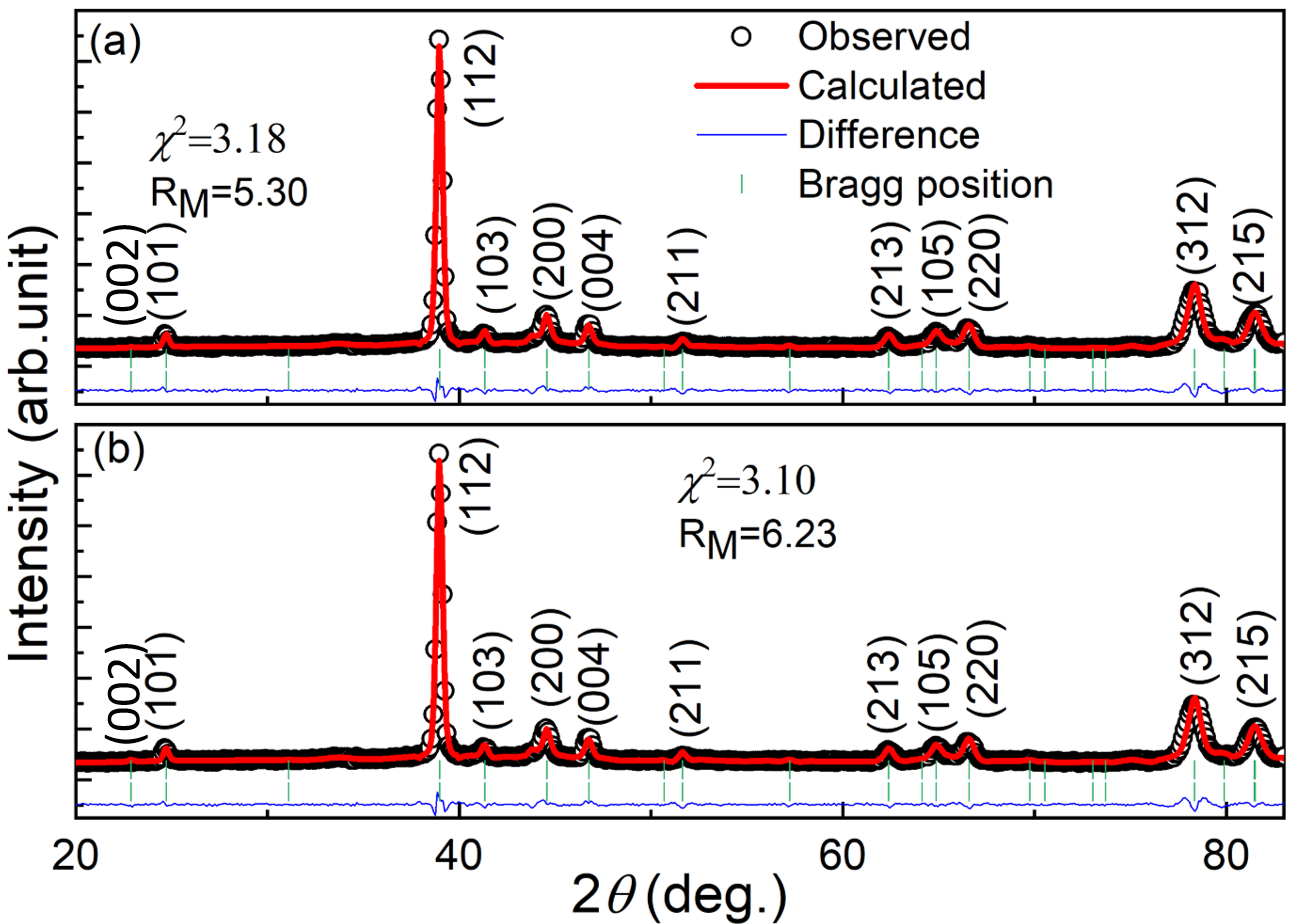}
		\caption{\label{FIG12} (Color online) Rietveld refinement of the neutron diffraction patterns at 1.5 K for Mn$ _{1.5} $PtIn for two different model (a) non-collinear and (b) non-coplanar. }
	\end{center}
\end{figure}

\begin{figure}[tb]
	\begin{center}
		\includegraphics[angle=0,width=8 cm,clip=true]{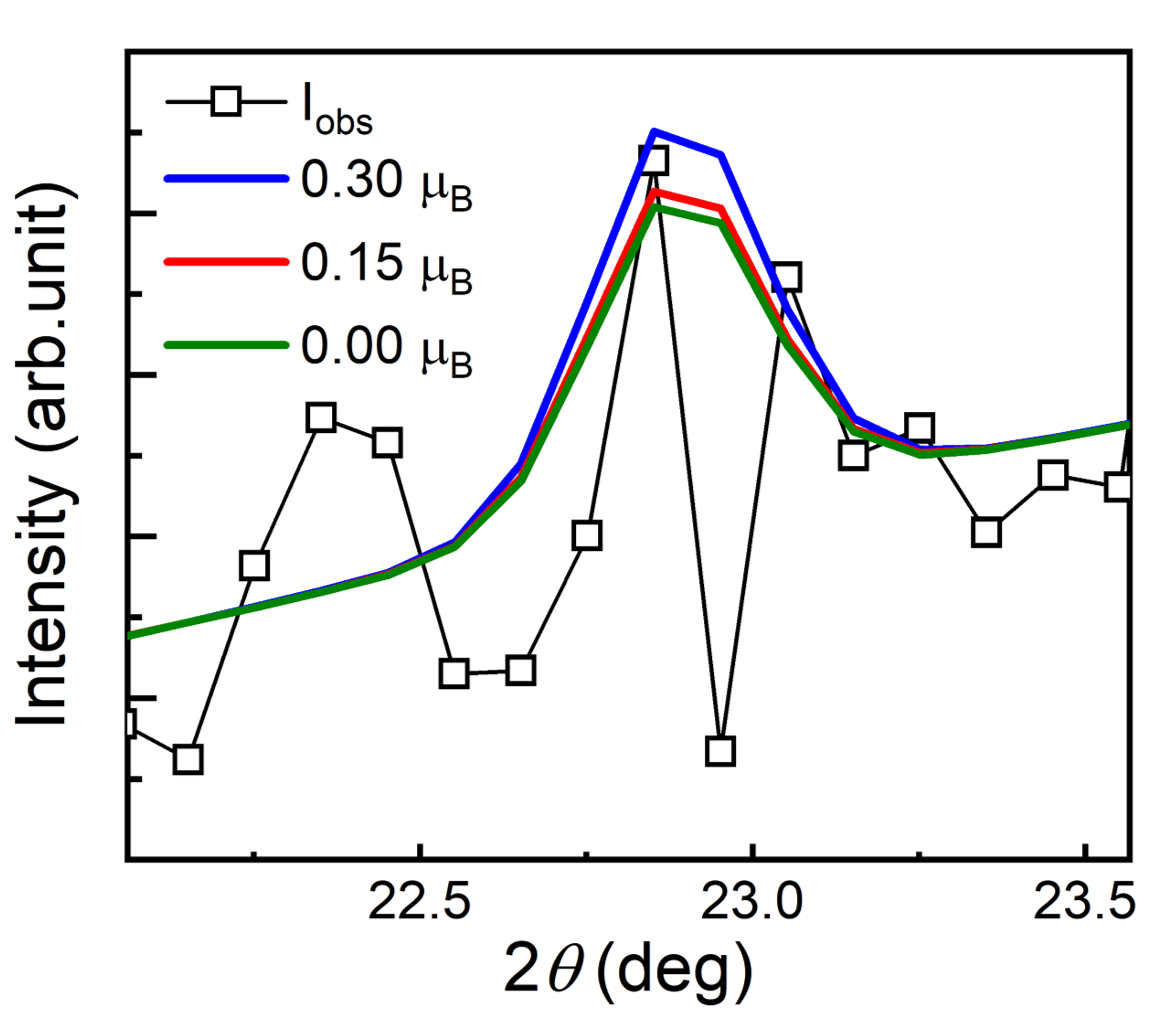}
		\caption{\label{FIG13} (Color online) Evolution of the calculated intensity on top of the (002) Bragg peak for the different magnitude of M$ _{X} $ component of Mn8i moment. }
	\end{center}
\end{figure}

\begin{figure}[tb]
	\begin{center}
		\includegraphics[angle=0,width=8 cm,clip=true]{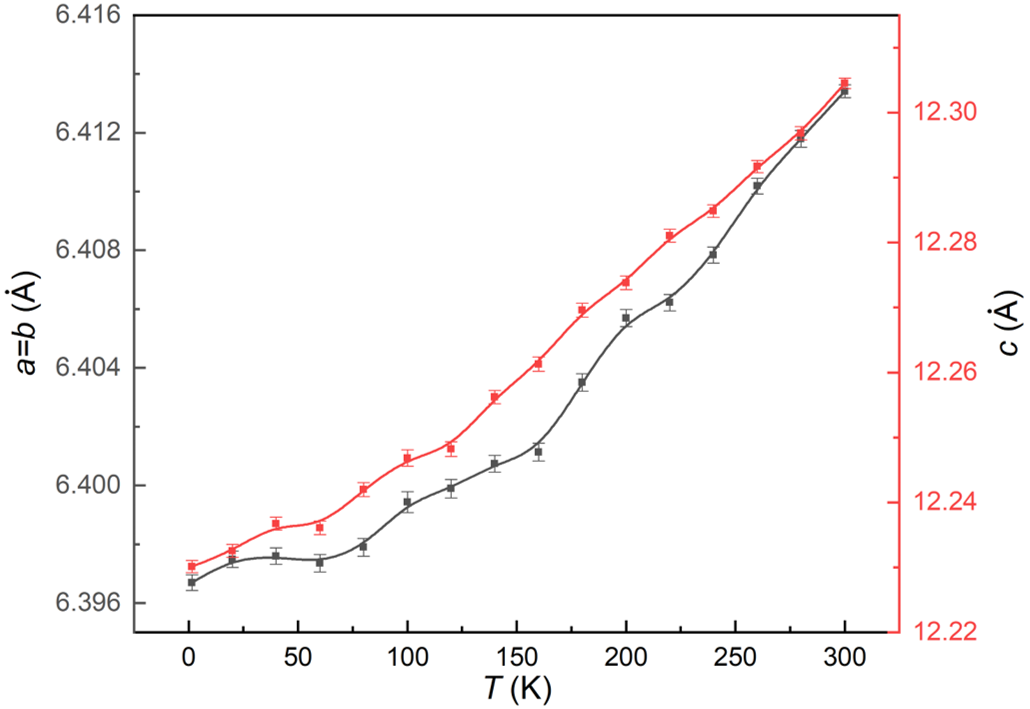}
		\caption{\label{FIG14} (Color online) Variation of lattice parameters with temperature of Mn$ _{1.5} $PtIn. The solid lines  are guide to eye. }
	\end{center}
\end{figure}

To determine the magnetic structure of the sample Mn$ _{1.5} $PtInn we performed the powder neutron diffraction (ND) at different temperatures. The ND patterns taken in the temperature range of 1.5 K to 300 K are depicted in the main manuscript. The intensity variation of the different Bragg’s peaks with temperature is shown in the main text. The temperature dependent ND pattern shows increase in the scattering intensity below the ordering temperature at the nuclear Bragg peaks (101), (200) and (004), suggesting a commensurate magnetic structure. Furthermore, we obtained the magnetic propagation vector k= (0, 0, 0) with best agreement factors by using the k-search programme included in Fullprof-suite package. We used the previously determined space group I\={4}2m (SG No. 121) and structural parameters for the Rietveld refinement. We have performed the Rietveld refinement of the 300 K ND pattern using only the nuclear contribution. The fit parameters obtained from the 300 K refinement were used for the refinement of structural part for the low temperature data. The Rietveld refinement of 1.5 K ND data convincingly demonstrates the presence of magnetic contribution in Mn$ _{1.5} $PtIn. The four magnetic atoms situated in the four different Wyckoff positions. It is difficult to have unique magnetic representations as the number of magnetic atoms is large. Therefore, we tried with different models obtained from symmetry analysis.
The symmetry analysis of the magnetic structure was conducted by the software SARAh\cite{supplement4}. Mn2a and Mn2b contained one 1-dim irreducible representations (IRs) $ \Gamma $2 one time and one 2 dim irreducible representations (IRs) $ \Gamma  $5 one time. Mn4d site with two 1-dim irreducible representations (IRs) $ \Gamma  $1 and $ \Gamma  $2 contained one time and one 2-dim irreducible representations (IRs) $ \Gamma  $5 contains two times. Mn8i site with two 1-dim irreducible representations (IRs) $ \Gamma  $1 and $ \Gamma  $3 contained one time and two 1-dim irreducible representations (IRs) $ \Gamma  $2 and $ \Gamma  $4 contained 2 times.  Also one 2 -dim irreducible representations (IRs) $ \Gamma  $5 contains 3 times. The model that best describe the magnetic structure are in agreement with the irreducible representations (IRs) $ \Gamma  $5 (Mn2a, Mn2b and Mn4d) while Mn8i is represented by the superposition of irreducible representations (IRs) $ \Gamma $5 and $ \Gamma  $2. The moment direction of Mn2a, Mn2b and Mn4d  within the a-b plane are not fixed, therefore we fixed them parallel to the crystallohraphic [010] direction. Restricting the MX (Mn8i) = 0, results in the non-collinear ferrimagnetic (we refer as model-1) ordering while the alternating MX(Mn8i) component (+x, -x) result into a non-coplanar structure (we refer as model-2)[Fig. 11]. 
In model- 1, Mn8i moment posses with the M$ _{Y} $ and M$ _{Z} $ component and for model-2 (non-oplanar structure) along with these two component additional M$ _{X} $(Mn8i) took alternate value (sequencing as +x, -x). The refinement, by considering these two model shows very similar agreement as can be seen from the Fig. 12. The evidence of the non-collinear magnetic ordering is indiated by the increases in the intensity of the (101) and (004) Bragg peaks. The enhancement in the intensity of these Bragg peaks is forbidden if all the magnetic moment contained only in the crystallpgraphic [001] direction. However, the increment in the intensity of the (004) peaks also hints a non-coplanar ordering in this system. Again, the emergence of the (002) peak would have constitute the non-coplanar structure (reffered as model- 2) as a more favourable. This fact an be understood from the evolution of the calculated intensity of the (002) peaks with the variation of the magnitude of the M$ _{X} $(Mn8i) component as shown in Fig. 13. However, as shown in Fig. 13., the non-coplanar component might yield a very weak reflection which may not be separated from the background signal. Furthermore, the parameters that were varied during the refinement are the  scale, zero-shift, the lattice constants (a, c), projections of the spin [M$ _{Y}  $(Mn2a, Mn2b, Mn4d), M$ _{Y} $ (Mn8i), M$ _{X} $ (Mn8i),  and M$ _{Z} $ (Mn8i)], and overall isotropic displacement (temperature) factor. Initially, we varied the parameter one by one and then releases some or all together at a time. The temperature dependence of absolute values of the magnetic moments for Mn sitting at different sublattices of Mn$ _{1.5} $PtIn are shown in the main mausript. The Mn sitting at 2b, 4d and 8i display almost equal magnitudes of magnetic moments with similar temperature dependance. A smaller magnitude of 2a Mn moment is due to the fact that 2a site is comparatively less occupied in the present sample. For non-coplanar magnetic structure, the Mn(8i) moment takes a alternate   M$ _{X} $ component and we obtained small M$ _{X} $(Mn8i)= 0.30(1) $ \mu $$ _{B} $. The variation of the lattice constants with the temperature obtained from the neutron diffraction measurements is shown in the Fig. 14.
\section{Hall Effect Measurement}
The Hall Effect experiments were performed with rectangular shape Hall bars. Five probe Hall measurements were carried on the Hall bar samples of Mn$ _{2} $PtIn, Mn$ _{1.5} $PtIn and Mn$ _{1.2} $PtIn. 
The total Hall resistivity generally consists of two terms in trivial ferromagnetic/ferri-magnetic samples and can be expressed as $ \rho_{yx} $ =$  \rho_{N} $ + $ \rho_{AH} $, where$  \rho_{N} $ and $ \rho_{AH} $ are the normal and anomalous Hall (AH) resistivity respectively. Normal Hall ($ \rho_{N} $ )  is given by $ \rho_{N} $=$ R_{0} $H, where $ R_{0 } $ is the normal Hall coefficient and related to the carriers density of the materials and H is the external magnetic field. However, the origin of anomalous Hall ($ \rho_{AH}  $) effect evolves with time since the discovery of it and can be classified into two mechanisms (i) intrinsic and (ii) extrinsic. The notion of Berry-phase curvatures comes under the intrinsic mechanism, whereas, the skew scattering and side-jump belongs to the extrinsic mechanism\cite{supplement5}.  
\begin{figure}[tb]
	\begin{center}
		\includegraphics[angle=0,width=8 cm,clip=true]{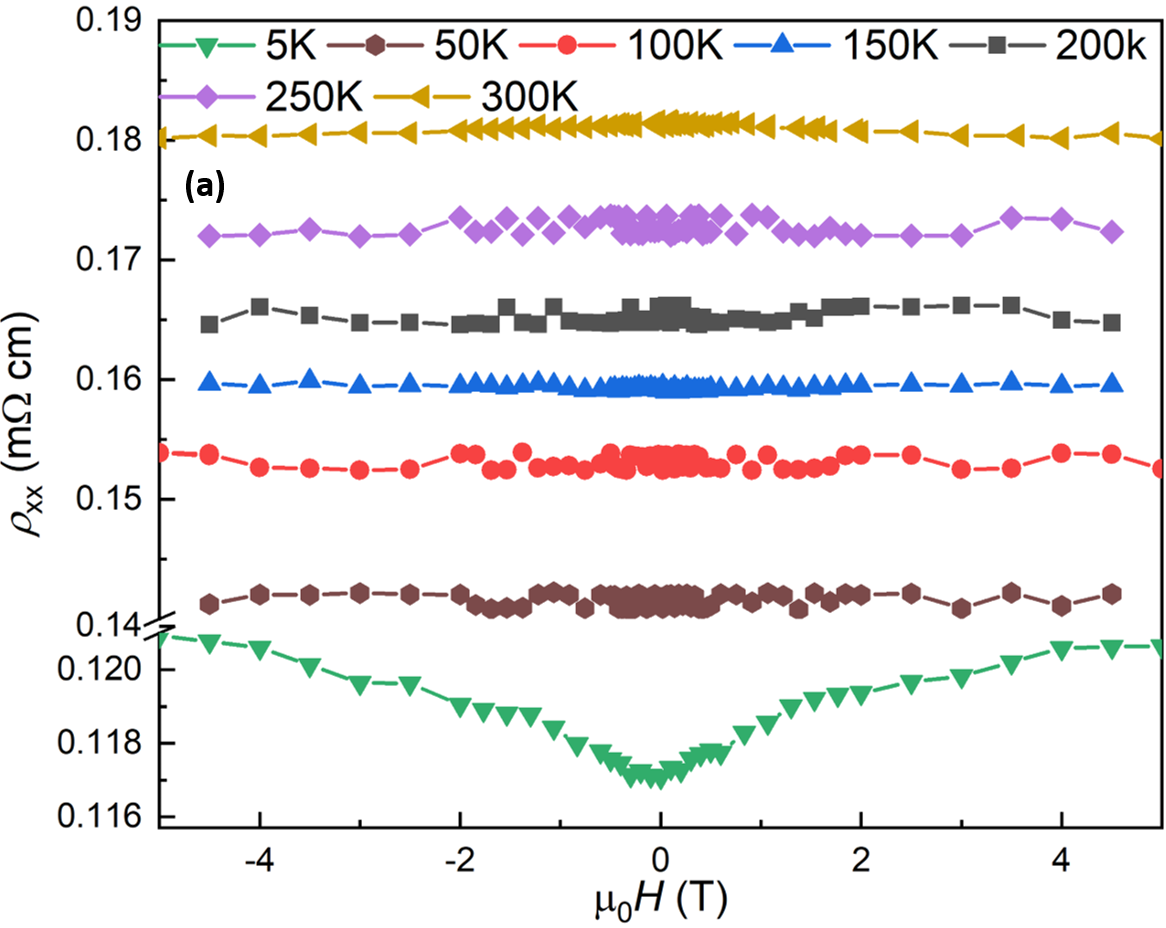}
		\caption{\label{FIG15} (Color online)  Field dependence of the longitudinal resistivity ($ \rho_{xx} $) for Mn$_{2}$PtIn at various temperatures. }
	\end{center}
\end{figure}
\begin{figure}[tb]
	\begin{center}
		\includegraphics[angle=0,width=8 cm,clip=true]{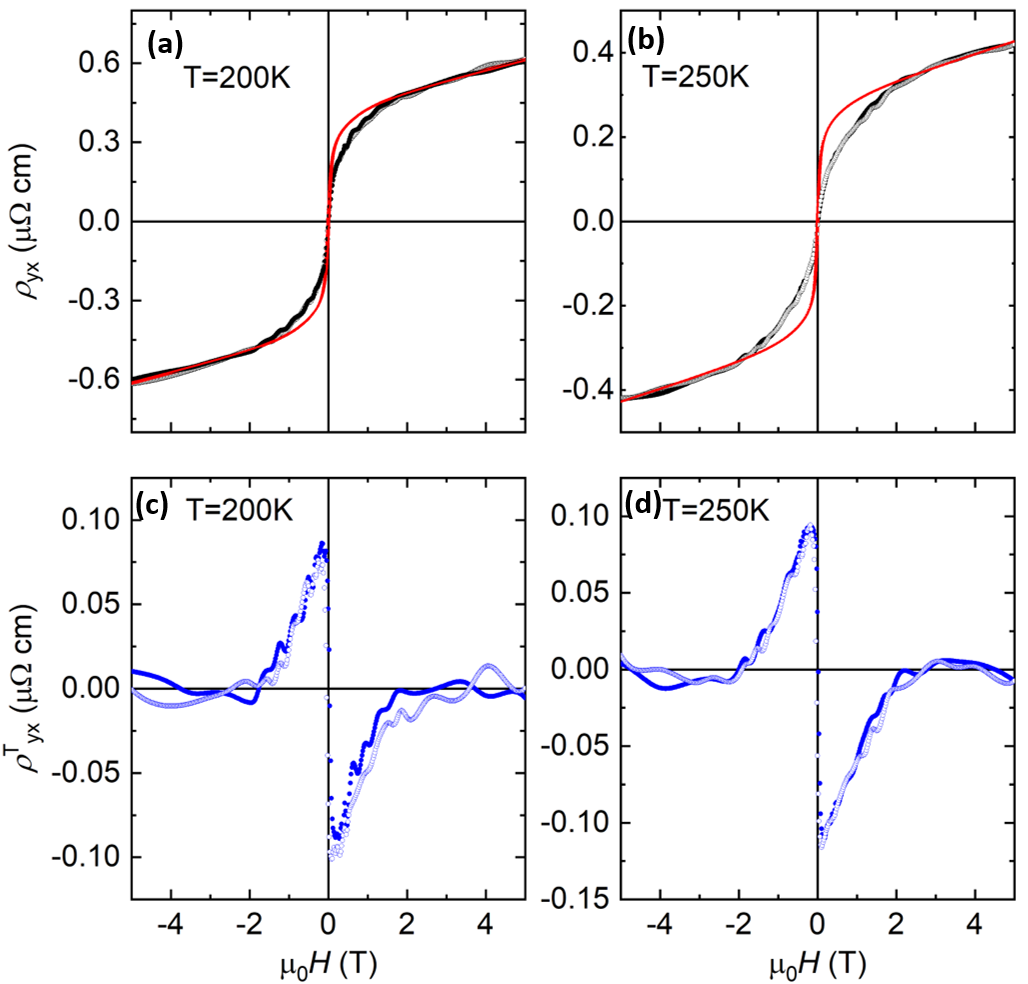}
		\caption{\label{FIG16} (Color online) Experimental and calculated Hall resistivity of Mn$_{2}$PtIn.  (a) and (b) experimental (black open (+5T to -5T) and solid (-5T to +5T) symbols) and calculated (solid red line) Hall resistivity curves at temperatures 200K and 250K respectively of the sample Mn$ _{2} $PtIn.  (c) and (d) corresponding to the topological Hall resistivity at different temperatures. Open symbols for +5T to -5T and solid symbols for -5T to +5T.}
	\end{center}
\end{figure}

\begin{figure}[tb]
	\begin{center}
		\includegraphics[angle=0,width=8 cm,clip=true]{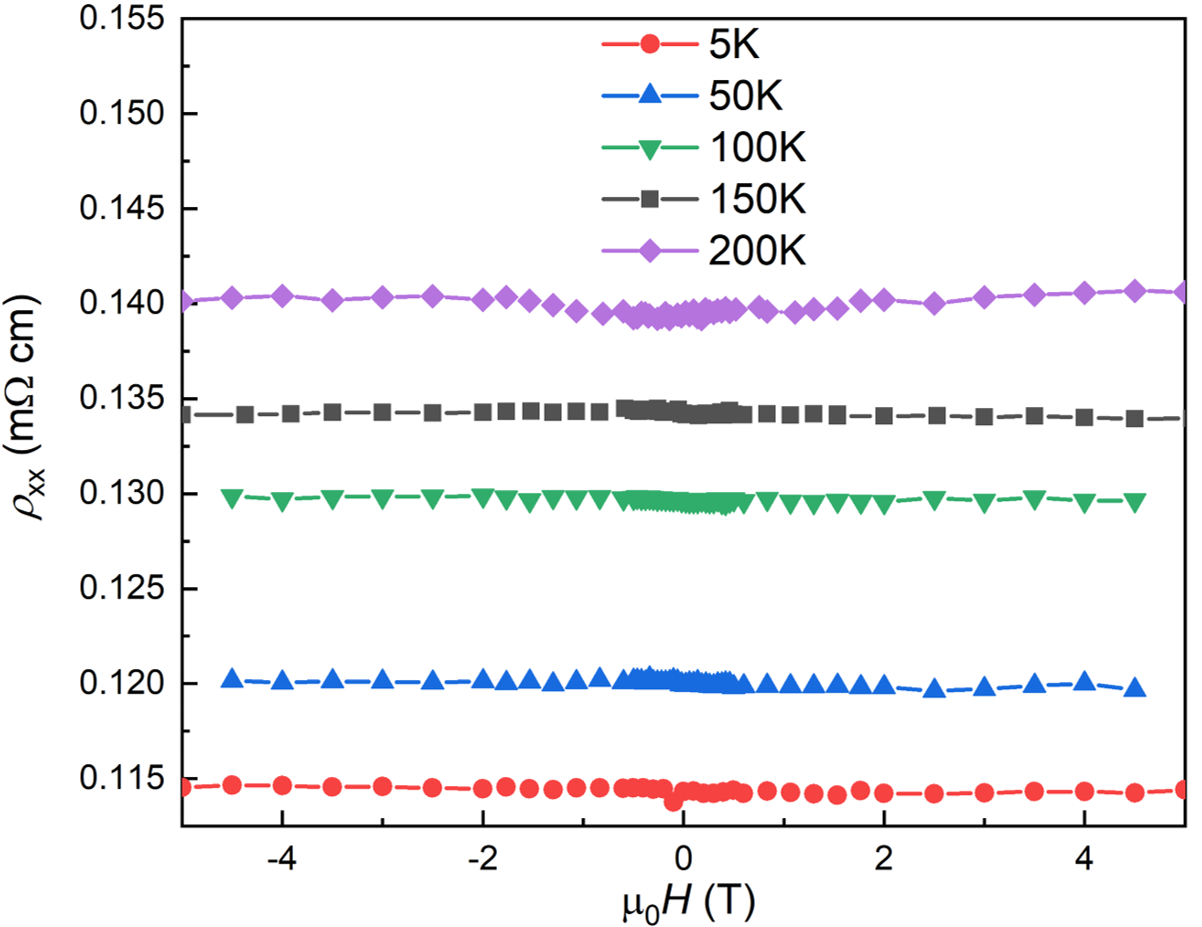}
		\caption{\label{FIG17} (Color online) Magnetic field dependence of the longitudinal resistivity ($ \rho_{xx} $) for Mn$_{1.5}$PtIn at various temperatures.}
	\end{center}
\end{figure}
\begin{figure}[tb]
	\begin{center}
		\includegraphics[angle=0,width=8 cm,clip=true]{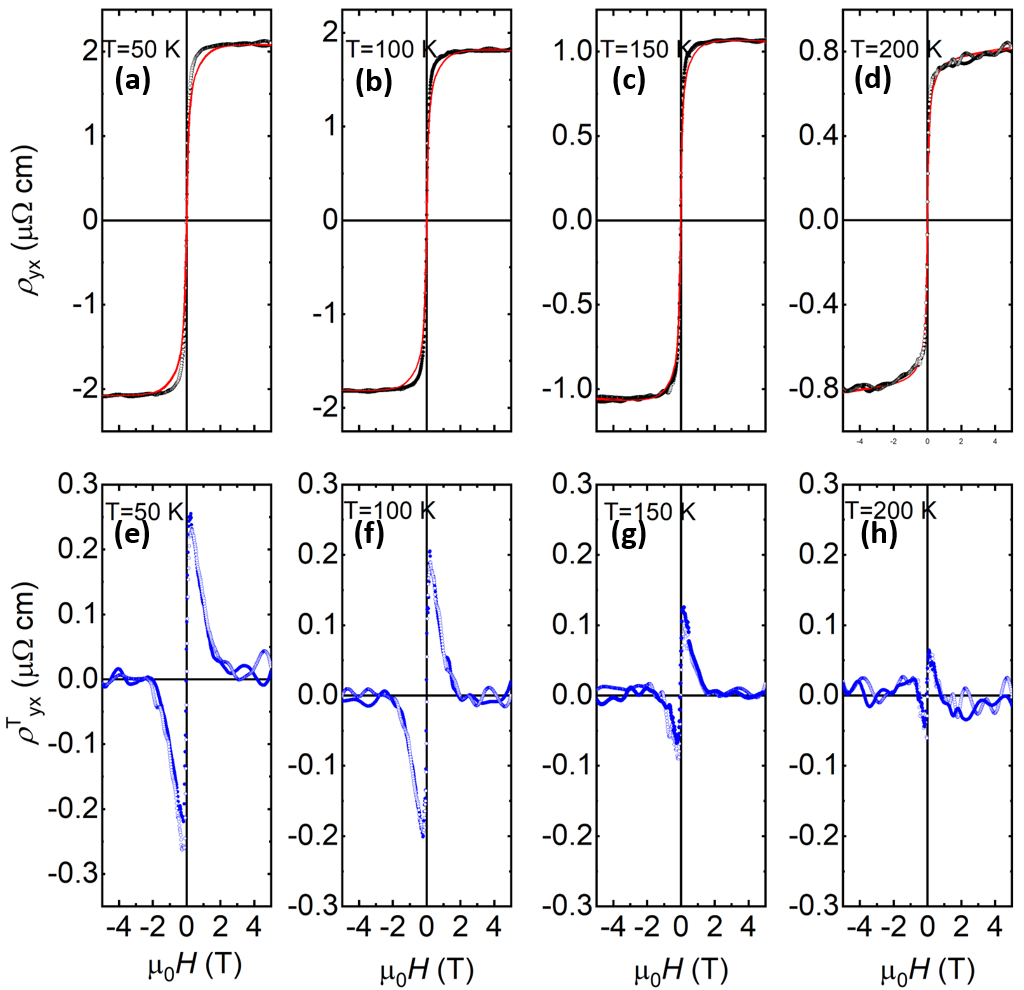}
		\caption{\label{FIG18} (Color online) Experimental and calculated Hall resistivity of Mn$_{1.5}$PtIn. (a), (b), (c) and (d) Experimental (black open (+5T to -5T) and solid (-5T to +5T) symbols) and calculated (solid red line) Hall resistivity curves at different temperatures for the sample Mn$ _{1.5} $PtIn.   (e), (f), (g), and (h) corresponding to the topological Hall resistivity at different temperatures.}
	\end{center}
\end{figure}

\begin{figure}[tb]
	\begin{center}
		\includegraphics[angle=0,width=8 cm,clip=true]{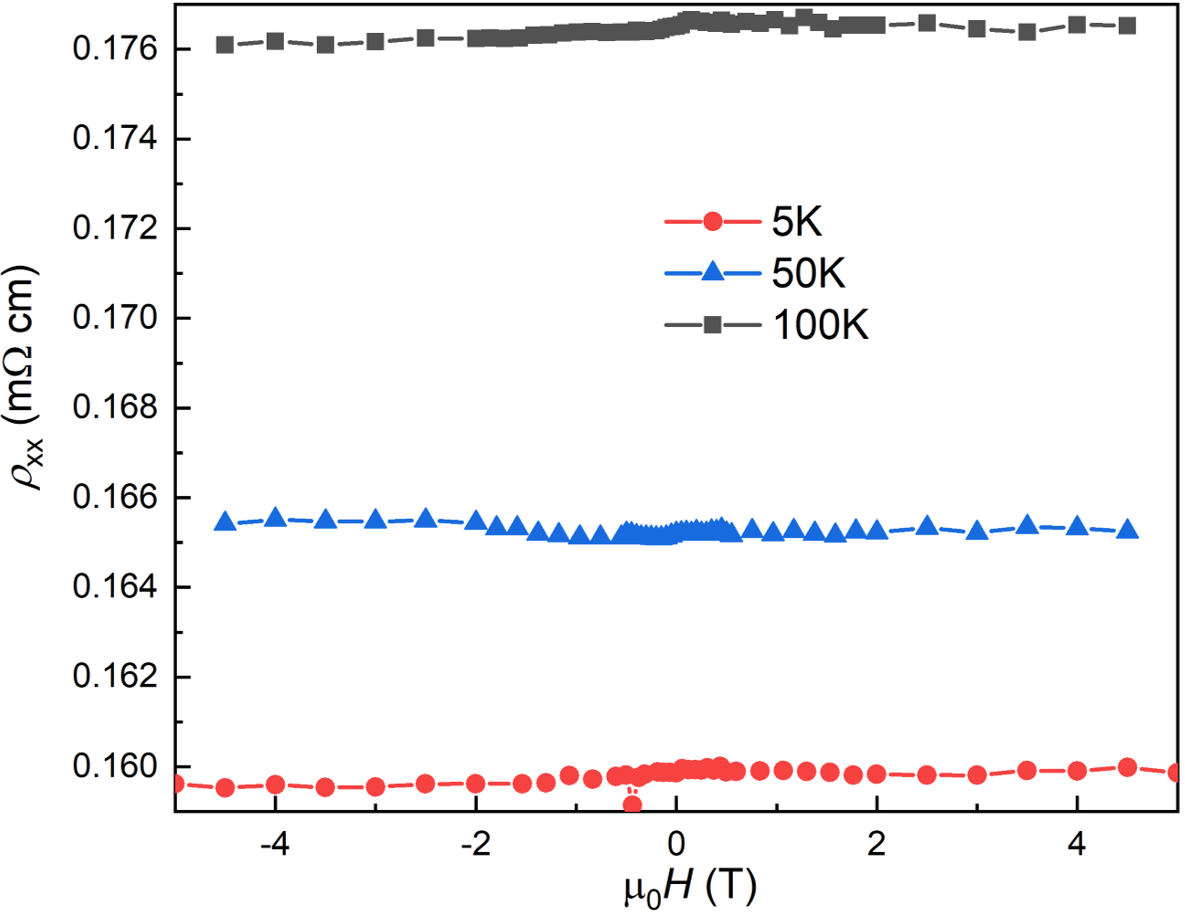}
		\caption{\label{FIG19} (Color online)  Field dependence of the longitudinal resistivity ($ \rho_{xx} $) for Mn$_{1.2}$PtIn at various temperatures.}
	\end{center}
\end{figure}
\begin{figure}[tb]
	\begin{center}
		\includegraphics[angle=0,width=8 cm,clip=true]{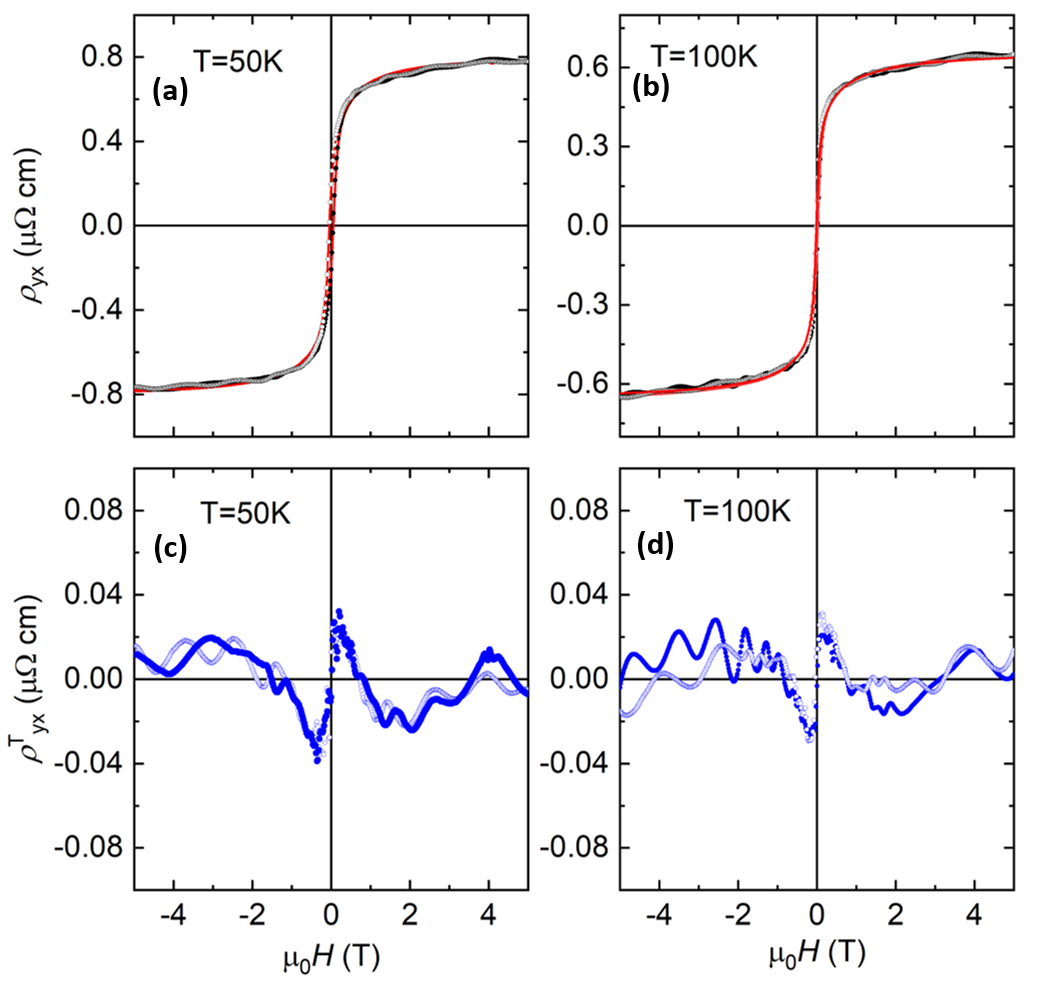}
		\caption{\label{FIG20} (Color online) Experimental and calculated Hall resistivity of Mn$_{1.2}$PtIn. (a) and (b) experimental (black open (+5T to -5T) and solid (-5T to +5T) symbols) and calculated (solid red line) Hall resistivity curves at different temperatures 50K and 100K respectively of the sample Mn$ _{1.2} $PtIn.  (c) and (d) corresponding to the topological Hall resistivity at different temperatures. Open and solid circles corresponding to +5T to -5T and -5T to +5T respectively.}
	\end{center}
\end{figure}
It is noteworthy to mention that the AH resistivity arise from the intrinsic mechanism varies as square of the longitudinal resistivity ($ \rho_{xx} $) and  directly scales with the magnetization (M) of the sample and can be written as $ \rho_{AH} $=b$\rho^2_{xx}$M, where b is the constant. However, the extrinsic contribution from the skew scattering ($\rho^{skew}_{AH}$) and side jump ($\rho^{sj}_{AH}$) to the AH resistivity can be express in terms of the longitudinal resistivity ($ \rho_{xx} $) such as $\rho^{skew}_{AH}$ $ \propto $ $ \rho_{xx} $  and $\rho^{sj}_{AH}$$ \propto $$\rho^2_{xx}$   \cite{supplement5}. Also, a new scaling relation demonstarted for $\rho^{skew}_{AH}$ and $\rho^{sj}_{AH}$ for high conductivity materials ($ \sigma_{xx} $ $\sim $$ 10^{6} $ $\Omega^ {-1} $ $ cm^{-1} $) in terms of the residual resistivity ($ \rho_{xx0} $) instead of $ \rho_{xx} $  \cite{supplement6}. Furthermore, different regime have been suggested to distinguish the dominant contribution from intrinsic or skew scattering process to AH resistivity based on the conductivity of the system under study \cite{supplement5}. The anomalous Hall effect due to skew scattering ($\rho^{skew}_{AH}$) and side jump ($\rho^{sj}_{AH}$) terms will be irrelevant even at low temperatures in materials with moderate conductivity ($ 10^{4} $ to $ 10^{6} $ $\Omega^ {-1} $ $ cm^{-1} $) with negligible magnetoresistance\cite{17,supplement7,supplement8,supplement9}. 
All the materials studied here falls under the moderate conductivity regime. So, $ \rho_{yx} $ can be written as  $ \rho_{yx} $= $ R_{0} $H + b$\rho^2_{xx}$M at the high field. The unknown constant$  R_{0}  $and b is obtained by a straight line ($ y $=m$ x $+c) fitting in the plot of [$ \rho_{yx} $/H versus ($\rho^2_{xx}$ M)/H]. The obtained parameters are utilized to calculate the complete Hall loop. As it can be seen from the Fig. 3 of the main manuscript that the calculated Hall resistivity matches perfectly with the experimental data at the high fields by considering only the intrinsic contribution to the anomalous Hall. The calculated Hall resistivity subtracted from the experimental Hall data to obtained the topological Hall contribution. Here it can be mentioned that taking only the intrinsic contribution in the anomalous Hall, the calculated and the experimental Hall resistivity data coinside with each other  for Mn$ _{1.2} $PtIn in all field regions. However, an adequate difference sustained for Mn$ _{2} $PtIn and Mn$ _{1.5} $PtIn in the low field regime. Figs. 15-20. show the field dependence of longitudinal resistivity and Hall resistivity for Mn$ _{2} $PtIn, Mn$ _{1.5} $PtIn and Mn$ _{1.2} $PtIn at various temperatures.


\end{document}